\documentstyle[prl,aps,amsfonts,amssymb,epsfig]{revtex}
\begin{document}
\draft 
\twocolumn[\hsize\textwidth\columnwidth\hsize\csname 
@twocolumnfalse\endcsname
\title{Infrared and optical spectroscopy of $\alpha$ and $\gamma$-phase Ce. }
\author{J.W. van der Eb, A. B. Kuz'menko\cite{byline}, and D. van der Marel}
\address{Material Science Center, University 
 of Groningen, Nijenborgh 4, 9747 AG Groningen, The Netherlands}
\date{\today}
\maketitle
\begin{abstract}
We determined the optical properties of $\alpha$- and $\gamma$-phase Ce in
the photon energy range from 60 meV to 2.5 eV using ellipsometry and
grazing incidence reflectometry. We observe significant changes of the
optical conductivity, the dynamical scattering rate, and the effective 
mass between $\alpha$- and $\gamma$-cerium. The $\alpha$-phase is
characterized by Fermi-liquid frequency dependent scattering rate, 
and a effective mass of about 20 m$_e$ on an energy scale of about 0.2 eV. In $\gamma$-Ce
the charge carriers have a large scattering rate in the far infrared, and a carrier
mass characteristic of $5d$ band electrons. In addition we observe a
prominent absorption feature in $\alpha$-Ce, which is absent in $\gamma$-Ce,
indicating significant differences of the electronic structure between
the two phases.
\end{abstract}

\vskip2pc]
\narrowtext
%
One of the most intriguing phenomena of cerium is the isostructural phase transition
between a low temperature $\alpha$-phase and a high temperature 
$\gamma$-phase, which ends in a solid-solid critical point, where the two phases are
indistinguishable. The structure is {\em fcc} with one atom per unit cell. It has become 
clear, that the proximity of the $4f$ level to the Fermi level is at the heart 
of these phenomena\cite{mcmahan}. Recently the nature of the $\gamma-\alpha$ phase transition 
has been questioned by Eliashberg and Capellmann\cite{eliashberg}. They argued
that a second order phase transition line continues beyond the tricritical
point on the melting curve, implying that
the symmetries of the two phases are different.  The question is
which symmetry breaking takes place through the phase transition.
Crystallographically the two phases appear to be identical\cite{xtalstruct}. 
Alternatively $\alpha$- and $\gamma$-Ce may
correspond to states of matter with the same 
crystallographic symmetries, but with a 'hidden' electronic symmetry which is 
broken in the $\alpha$-phase. This motivated us to embark on a series of 
spectroscopic experiments on the electronic structure\cite{jeb}
including the work presented here. Although we can not argue on the basis of
these experiments that a symmetry is broken, we do provide compelling evidence
that the electronic structure of both phases is qualitatively different.  

Although an extensive literature exists on optical properties of heavy fermion 
compounds, many of which are intermetallic compounds based on Ce\cite{leo}, 
little is known about the optical spectra of pure metallic cerium
\cite{wilkins,knyazev,rhee}. Wilkins {\em et al.} reported several absorption
peaks (288, 218, 170, 150, 83  $meV$)\cite{wilkins}, which we have not been
able to reproduce. Rhee {\em et al.} reported optical spectra from 1.5 to 5 eV
on different films\cite{rhee}, with a large sample dependence of the data, 
and which were distinctly different from ellipsometry measurements 
of polycrystalline samples by  Knyazev {\em et al.}\cite{knyazev}. 
In this Letter
we present infrared/optical data over a broad range of photon energy. 
Using {\em in-situ} thin film deposition technology we obtained excellent 
sample to sample reproducibility of the optical spectra.

Cerium is a highly reactive metal. In order to prepare and maintain a clean and oxygen
free sample over a period of about one day, it is necessary
to work under ultra high vacuum. We used a compact
UHV cryostat for low temperature {\em in situ} evaporation of Ce films, 
with vacuum in the $10^{-10}$ $mbar$ range. The 
ZnSe windows are transparent from 0.06 to 2.5 eV photon energy.
Following the procedure outlined in Ref.\cite{films} we
prepared thin films of $\alpha$- and $\gamma$-phase Ce on clean and oxygen free silicon 
wafers using ultra high vacuum deposition. The evolution of the optical spectra 
was monitored during the deposition process. The growth rate of about $2\AA/$s
provided smooth mirror-like films, in agreement with earlier reports\cite{films}, 
with optical properties which were reproducible over 4 cycles of deposition and 
measurement. Details of the sample preparation are given in Ref. \cite{jeb}. 

\begin{figure}
    \centerline{\psfig{figure=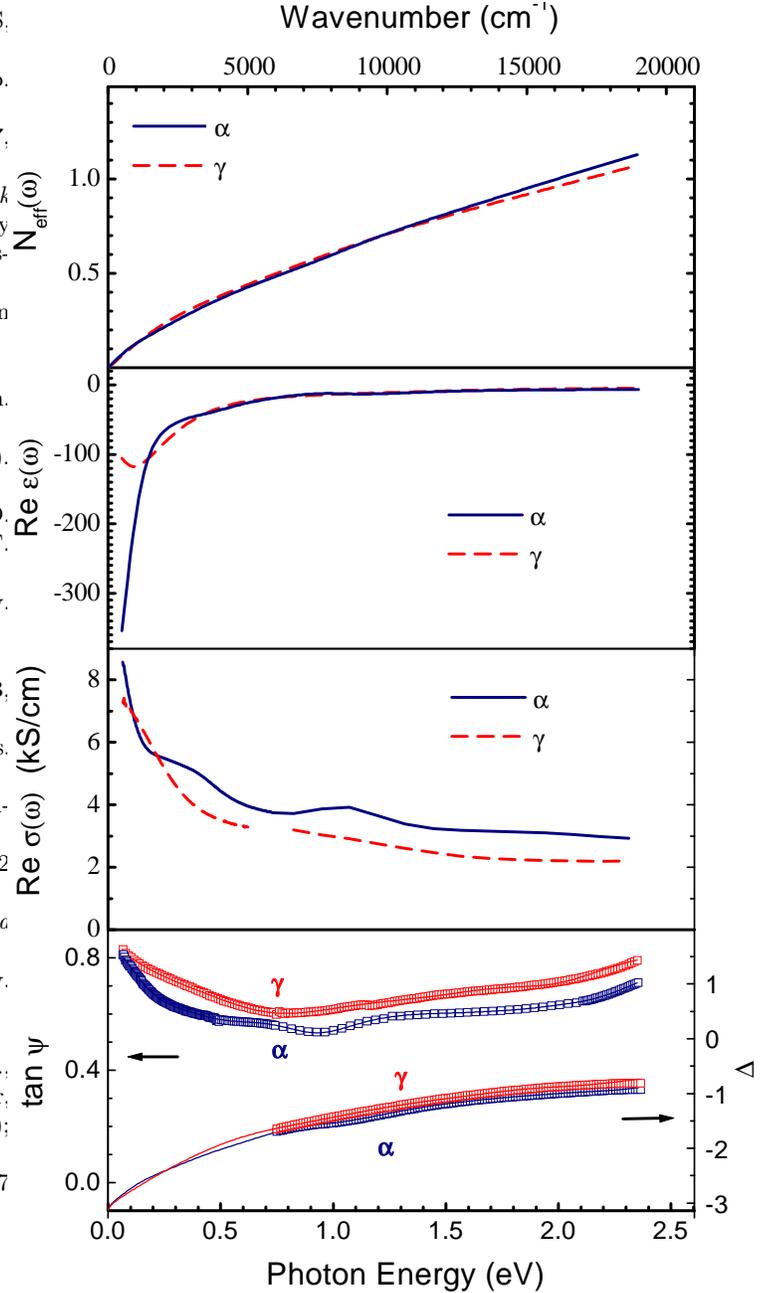,width=10cm,clip=}}
    \caption{ $\psi(\omega) $ and $\Delta(\omega)$. Open symbols: 
             Measured data. The solid curves are calculated using the
             analysis explained in the text, and they overlap exactly
             with the symbols.
             Real part of the optical conductivity (second panel from below),
             real part of the dielectric-constant (third panel from below)
             and spectral weight function (top panel)
             of $\alpha$- and $\gamma$-Ce. }
     \label{epssig}
\end{figure}

For photon energies between 0.8 and 2.5 eV we measured the optical 
constants using spectroscopic ellipsometry using an angle of incidence 
$\theta$ of 80$^{o}$. Ellipsometry
provides directly the amplitude-ratio ($\tan\psi=|r_p/r_s|$) ) and phase-difference
($\Delta= \mbox{Arg}(r_p)-\mbox{Arg}(r_s)$) ) 
of the reflection coefficients ($r_p$ and $r_s$) of 
$p$- and $s$-polarized light, the dependence of which on the dielectric function,
$\epsilon(\omega)$ is given by the ellipsometric formula
\begin{equation}
 \tan \psi e^{i\Delta} = \frac{\sin^2\theta-\cos\theta\sqrt{\epsilon-\sin^2\theta}}
                         {\sin^2\theta+\cos\theta\sqrt{\epsilon-\sin^2\theta}}
\label{ellips}
\end{equation}
In the photon energy range of 60 meV to 0.8 eV we used a Fourier transform 
infrared spectrometer to measure the intensity ratio of the
reflectivity coefficients of $p$- and $s$-polarized light, providing
only $\psi(\omega)$, but not $\Delta(\omega)$. The 
conventional approach for obtaining $\Delta(\omega)$ would be, to 
use Kramers-Kronig relations between phase
and amplitude. However, knowledge of the amplitude 
over only a limited frequency range is in principle insufficient to determine 
the phase function. As was shown in Refs.\cite{milton,bozovic} a minimum of two 
'anchors' (frequencies where amplitude {\em and} phase have been measured) suffices to
nail down the phase in the entire frequency range where the amplitudes have been
measured. In the present case we have anchored the phases from 0.8 eV to 2.5 eV, 
which allows us to determine the $\Delta(\omega)$ between 
0.06 and 0.8 eV in a very accurate way. Special care is needed here, because
for $\theta > 45^0$ the function 
$\ln(\tan\psi(\omega))+i\Delta(\omega)$ does not 
have all its poles in the lower half of the complex frequency-plane, as
can be verified easily from Eq. \ref{ellips}. 
On the other hand, Re$\epsilon(\omega)$ and 
Im$\epsilon(\omega)$ {\em do} satisfy this condition, implying that
$\epsilon(\omega)$ can always be represented as a sum of oscillators
$\epsilon(\omega)= 1 + \sum_j
\omega_{pj}^2\{\omega_{0j}^2-\omega^2-{\rm i}\gamma_j\omega\}^{-1}$
This means that any experimental $\psi(\omega)$ and/or $\Delta(\omega)$ 
spectrum can be fitted by adjusting the number of oscillators such, as to reproduce
all the details of the spectrum other than statistical noise. No obvious 
physical interpretation of the individual oscillators
necessarily exists, as the oscillator sum is rather
similar in nature to a Kramers-Kronig integral, which indeed also becomes a 
summation when applied to a set of experimental data points. 
Using Eq. \ref{ellips}, we determined $\Delta(\omega)$ for $\hbar\omega < 0.8$ eV 
with this analysis, using on input the experimental $\psi(\omega)$
for 0.06 eV $< \hbar\omega < 2.5$ eV, and $\Delta(\omega)$ for 
$\hbar\omega >  0.8$ eV (see bottom panel of Fig. \ref{epssig}). The final 
step is to calculate Re$\epsilon(\omega)$ and 
Re$\sigma(\omega)=\omega$Im$\epsilon(\omega)/4\pi$
from the ellipsometric parameters using  Eq. \ref{ellips}. For
$\omega > 0.8$ eV, we used the directly measured $\psi$ and $\Delta$.
For $\omega < 0.8$ we used the directly measured $\psi(\omega)$ and 
$\Delta(\omega)$ determined with the analysis described above.

Following the analysis of Ref.\cite{films} we
identify the films deposited at temperatures below 4 K and above 300 K 
($\sim$400 K in our case) as $\alpha$- and $\gamma$-phase 
Ce respectively. We observed strong hysteresis of the optical spectra during
cycling of the temperature after deposition 
Similar hysteresis loops have been reported for the DC resistivity 
of Ce films by L\"offler {\em et al.}\cite{loffler}.
In Fig. \ref{epssig} we show Re$\epsilon(\omega)$ and Re$\sigma(\omega)$ 
for the $\alpha$- and the $\gamma$-phase.
In the $\alpha$-phase the Drude contribution narrows and increases in
intensity, leading to a more metallic-like behavior as expected.

At the lowest frequencies of our experiment
the conductivity of $\alpha$-phase cerium is
higher than that of the $\gamma$-phase, in qualitative agreement with the
trends observed in the DC transport properties\cite{loffler}.
In addition to a very sharp Drude peak, we observe in $\alpha$-Ce a shoulder
at 0.3 - 0.4 eV, and a peak at 1.0 $eV$, on top of a broad and featureless
background conductivity. In the $\gamma$-phase the Drude peak is much
broader, the shoulder is absent, and again there is a broad and almost
featureless background, except for a weak and broad shoulder from 0.5 to
1.5 eV.  The most significant difference is the presence of a prominent 
peak at 1 eV in the $\alpha$-phase, and the absence thereof in the 
$\gamma$-phase, indicating that the electronic structure of both phases
is significantly different on an energy scale far exceeding reported values 
of the Kondo temperature. The intensity and sharpness of 
this peak suggests that this is a virtual bound state of 
$5d\rightarrow 4f$ character.

Cerium has $4$ valence electrons per unit cell. One of the key issues is how these
electrons are distributed between the quasi-atomic $4f$-states and the non-$4f$
conduction bands (mainly $5d$), and to what extent this distribution changes
as a function of temperature. A powerful tool to address these issues is
provided by an analysis of the spectral weight. In the top panel of
Fig.\ref{epssig} we display the function
\begin{equation}
 N_{eff}(\omega)=\frac{2m_e\Omega_u}{\pi e^2 }\int_0^{\omega}\sigma(\omega')d\omega'
\end{equation}
where we used $\Omega_u=29.2 \AA^3$ for the unit cell volume of
$\alpha$-Ce and $\Omega_u = 34.4 \AA^3$ for $\gamma$-Ce.
According to the $f-$sum rule the limiting behaviour at high
frequencies $N_{eff}(\omega)$ reflects the {\em total}
number of electrons (core and valence). If the 
integration is restricted to the Drude peak,  $N_{eff}(\omega)$ corresponds to 
the $k$-space (Luttinger) volume occupied by the valence electrons, multiplied
by $m_e/m_b$, where $m_b$ is the band-mass of the conduction electrons.
For both phases $N_{eff}(\omega) \sim 1$ between 1.5 and 2.5 eV. 
However, even at 2.5 eV there is no saturation of $N_{eff}(\omega)$.
This implies, that interband transitions and electron-electron scattering
processes continue to contribute to the optical absorption in the ultraviolet.
An upper limit of $m_b$ can be set to
$\sim 6 m_e$, which is reasonable given the fact that these
bands have mainly $5d$ character\cite{pickett,liechtenstein}.

Let us consider the case of $\alpha$ cerium at low frequencies, and let us assume 
at this point that the narrow Drude peak 
below $\sim 0.2$ eV is the response of all four valence electrons, 
with the same effective mass. To estimate the spectral weight
of the narrow peak, we choose 
$\omega \approx 0.2 eV$ as the upper limit of integration. 
Because $N_{eff}(0.2 eV)\approx 0.2$, this implies
that close to E$_F$ the valence electrons have an effective mass
$m^*/m_e\approx 20$. In reality the Fermi surface has a complicated shape,
and the mass may have a strong $k$-dependence along the various sections of
this Fermi surface. As a consequence $m^*$ should be regarded as a quantity
representing the $\em average$ effective mass.

Let us now consider the $\gamma$-phase. In this case we have a good reason to
make a clean distinction between the localized $4f$ states and the conduction
bands of mainly $5d$ character. Due to their localized character the $4f$ electrons
have a negligible contribution to the low frequency spectral weight. As a result
the Drude peak extending upto about 1 eV is mainly of $5d$ character, and 3 electrons
contribute to the Drude peak. Reading $N_{eff}(0.6 eV)\approx 0.5$ for
the $\gamma$-phase, we conclude that the effective mass of the conduction
electrons of $\gamma$-Ce is about $\sim 6 m_e$.

\begin{figure}
    \centerline{\psfig{figure=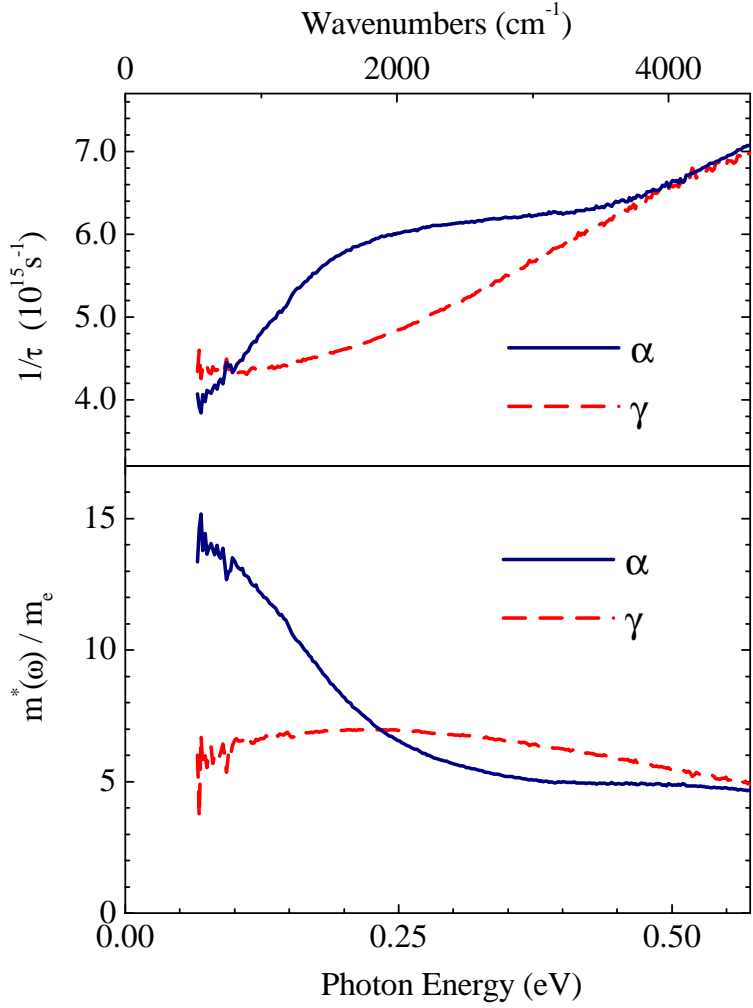,width=10cm,clip=}}
    \caption{Frequency-dependent effecive mass (bottom panel) and scattering
             rate (top panel) of $\alpha$- and $\gamma$-Ce. } 
    \label{mtau}
\end{figure}

Another way to analyze the nature of the low frequency spectral weight is
to calculate the frequency-dependent effective mass $m^*(\omega)$ and scattering
rate $1/\tau(\omega)$ directly from the real and imaginary part of the
optical conductivity, using the generalized Drude form \cite{jallen1}
\begin{equation}
 \sigma(\omega)=\frac{ne^2/m_e}{ \tau(\omega)^{-1} + i \omega m^*(\omega) / m_e}
\end{equation}
using $n=4/\Omega_u$ for the density of valence electrons.
The results are shown in Fig.\ref{mtau}. 
The large mass enhancement in the $\alpha$-phase indicates that the valence
bands at the Fermi energy are strongly renormalized. The 
heavy mass indicates that at least one of the bands crossing the
Fermi has considerable $4f$ character, and/or a strong mass-renormalization due
to coupling to spin-fluctuations. 
The width of the effective mass peak is about $0.2$ eV, which indicates that
coherent Bloch like states exist below a typical energy scale 
$k_B T_{coh}=0.2$ eV\cite{fulde}.
At higher energies the Fermi-liquid type mass renormalization reduces
gradually to the mass of a bare band electron, and
in Fig. \ref{mtau} we see, that the effective mass of the $\alpha$-phase
approaches that of the $\gamma$-phase above 0.6 eV. 
The frequency dependent scattering rate of the $\alpha$-phase bears a 
strong resemblance to Fig. 13.3 of Ref.\cite{fulde}, which in the
context of the Kondo lattice model, indicates that a narrow band of itinerant
electrons is formed with a coherence temperature $k_B T_{coh} = 0.2$ eV, and
a characteristic heavy-fermion temperature $k_B T^* = 0.4$ eV. 

The frequency dependence of the scattering rate and effective mass of
$\alpha$-Ce has similar characteristics
as the heavy fermion compound URu$_2$Si$_2$, for
which $k_B T_{coh} = 6$ meV and a mass enhancement factor 
of 50 have been reported\cite{bonn}. The frequency dependence $1/\tau$
and $m^*$ (Fig. \ref{mtau} are directly visible  
as a shoulder at 0.3 eV in the conductivity of $\alpha$-Ce (Fig. \ref{ellips}). 
The spectra of $\alpha$-and $\gamma$-Ce agree very well with calculations
based on the Local Impurity Self Consistent Approximation of the Anderson model
by Rozenberg, Kotliar, and Kajueter (Fig. 20 of Ref. \cite{rozenberg}) for temperatures
far below ($\alpha$-C) and above ($\gamma$-Ce) $T^*$. Similar side-bands have
been observed in UPd$_2$Al$_3$ (1 meV\cite{dressel}), UPt$_3$ (1 meV)\cite{gruner}),
CeCu$_6$(5 meV\cite{marabelli}), UNi$_2$Al$_3$ (25meV\cite{cao}), and 
Yb$_{1-x}$Lu$_x$B$_{12}$($(1-x)\cdot$0.25 eV\cite{okamura}). It is conceivable
that at least for some of these other materials the side band is of a similar
nature as in $\alpha$-Ce.  The large variation of the position of the side-band
among the different heavy fermion systems would then reflect a variation
of the coherence temperature from system to system, which is a lower 
energy scale than the on-site electron-electron interactions (Hubbard $U$)\cite{rozenberg}.  
The hierarchy of energy scales in heavy-fermion systems is believed to be 
$T_{coh}\le T^* \le T_K$, providing indirect evidence that $T_K$ is several
thousend K for $\alpha$-Ce, in agreement with the analysis of Allen and 
Martin\cite{jallen2}. They estimated, that $T_K$
of the $\gamma$-phase is of order 20 K \cite{jallen2,gs}. Under these 
conditions the mass renormalization effects are small. On the other hand 
the partly filled $4f^1$ states at each Ce atom still 
act as paramagnetic impurities, which form an efficient source for 
resonant scattering of the valence electrons. This is visible
as a large residual scattering rate, as suggested by saturation of 
$1/\tau(\omega)$ to a high value in the far infrared range. 

An interesting consequence emerges from these considerations:
$\gamma$-Ce corresponds to a state of matter with 3 valence electrons per atom.
The fourth electron only contributes a single spin degree of freedom,
while the charge degree of freedom is gapped at an energy scale of several eV.
The $\alpha$-phase on the other hand, corresponds to a state where
all 4 valence electrons participate in itinerant bands. This result has
common factors with the Mott model of Johansson\cite{johansson} and with 
the Kondo Volume Collaps model\cite{jallen2,gs}. The difference with
the Mott model is, that not four, but only one $4f$ electron undergoes a Mott
transition. With the KVC model it shares the property that the $\gamma$-phase 
consists of a band of itinerant electrons, weakly exchange coupled to 
paramagnetic local moments. The main difference is for the $\alpha$-phase, where
our optical data suggest that an itinerant band is formed, a feature which
requires explicit consideration of the lattice aspect of the problem. These
aspects have been recently taken into account by Laegsgaard and Svane\cite{svane}
in a model combining SIC-LDA and the Anderson impurity model. 
Is indeed a hidden electronic symmetry broken in the $\alpha-\gamma$ transition? 
The disappearance of one charge degree of freedom per atom
as the system enters the $\gamma$-phase, suggests that the answer to this 
question is affirmative.

We performed infrared grazing incidence measurements on cerium thin films. 
The conductivity shows a remarkable change between the $\alpha$ and $\gamma$
phase.  A significant difference is the presence of a prominent peak at 1 eV
in the $\alpha$-phase, and the absence thereof in the $\gamma$-phase, indicating
that the electronic structure is significantly different in both phases on
an energy scale far exceeding reported values of the Kondo temperature of
both phases. 
At low frequencies $\alpha$-Ce is found to be reminiscent of a Fermi-liquid 
with an effective mass of $20m_e$. We observe this behaviour on an energy scale 
of about $k_B T_{coh} = 0.2$ eV, above which $\tau(\omega)^{-1}$ and $m^*(\omega)$ 
saturate at $6\cdot 10^{15}$s$^{-1}$ and $5m_e$ respectively. 
In the $\gamma$-phase the three conduction electrons have an effective 
mass of $\sim 6 m_e$, with a large residual scattering rate ($4\cdot 10^{15}$s$^{-1}$)
in agreement with the notion of valence electrons  resonantly scattered by 
$4f^{1}$ local moments.

We like to thank G. M. Eliashberg for drawing our attention to this problem.
A.B.K. was supported by the Russian Foundation of Basic Research, grant No
99-02-17752 and by the Nederlandse Organisatie voor Wetenschappelijk Onderzoek
(NWO). This investigation was supported 
by the Netherlands Foundation for Fundamental Research on Matter 
(FOM) with financial aid from the Nederlandse Organisatie voor 
Wetenschappelijk Onderzoek (NWO).

\newpage
\newpage


\begin{references}
\bibitem[*]{byline}Also: P.L.Kapitza Institute for Physical Problems RAS,
Kosygina str. 2, Moscow, 117334, Russia
\bibitem{mcmahan}
  For a recent review, see A. K. McMahan {\em et al.}, J. Comp. Mat. Design {\bf 5}, 131 (1998).
\bibitem{eliashberg}
  G. M. Eliashberg, and H. Capellmann, JETP Lett. {\bf 67}, 125 (1998).
\bibitem{xtalstruct}
  D. A. Koskenmaki and K. A. Gschneidner, in {\em Handbook on the Physics and
  Chemistry of Rare Earths}, edited by K. A. Gschneidner and L. Eyring, North-Holland,
  Amsterdam, 1978, Chap. 4.
\bibitem{jeb} 
  J.W. van der Eb, PhD thesis, University of Groningen (2000).
\bibitem{leo}
  L. Degiorgi, Rev. Mod. Phys. {\bf 71}, 687 (1999).
\bibitem{wilkins} 
 J.F. Wilkins, J.G. Clark, T.E. Leinhardt, Bull. Am. Phys. Soc. {\bf51}, 579 (1962).
\bibitem{knyazev} 
  Yu.V. Knyazev {\em et al.},
  Phys. Nizk. Temp. {\bf17}, 1143 (1991).
\bibitem{rhee} 
  J.Y. Rhee {\em et al.},
  Phys. Rev. B {\bf51}, 17390 (1995).
\bibitem{films}
  E. Weschke {\em et al.}, Phys. Rev. B {\bf 44},  8304  (1991);
  D. Wieliczka {\em et al.},  Phys. Rev. B {\bf 26}, 7056 (1982);
  F. Patthey {\em et al.},  Phys. Rev. Lett. {\bf 55}, 1518  (1985).
\bibitem{milton}
  G. W  Milton, D. J. Eyre, and J.V. Mantese, Phys. Rev. Lett. {\bf 79}, 3062 (1997).
\bibitem{bozovic}
  I. Bozovic, Phys. Rev. B {\bf 42}, 1969 (1990).
\bibitem{loffler} 
  E. L\"offler, and J.A. Mydosh, Solid State Commun. {\bf13}, 615 (1973).
\bibitem{pickett}
  W.E.Pickett, A.J. Freeman, and D. D. Koelling, Phys. Rev. B {\bf 23}, 369 (1981).
\bibitem{liechtenstein}
  A. B. Shick, W. E. Pickett, and A. I. Liechtenstein, cond-mat/0001255 
\bibitem{jallen1} 
  J. W. Allen,  and J.C. Mikkelsen, Phys. Rev. B {\bf15}, 2952 (1977).
\bibitem{fulde}
  P. Fulde, in {\em Electron Correlations in Molecules and Solids}, third edition,
  Springer, Berlin, 1995. 
\bibitem{bonn}
 D. A. Bonn, J. D. Garett, and T. Timusk, Phys. Rev. Lett. {\bf 61}, 1305 (1988).
\bibitem{dressel}
 M. Dressel {\em et al.}, Physica B {\bf 244} 125 (1998).
\bibitem{gruner}
 G. Gr\"uner, Physica B {\bf 244} 70 (1998); 
 P.E. Sulevski {\em et al.}, Phys. Rev. B {\bf 38} 5338 (1988);
 F. Marabelli, P. Wachter, and J.J.M.Franse, J. Magn. Magn. Mater. {\bf 62}, 287 (1986);
 A. Awasthi {\em et al.}, Phys. Rev. B {\bf 39}, 2377 (1996).
\bibitem{marabelli}
 F. Marabelli, and P. Wachter, Phys. Rev. B {\bf 42}, 3307 (1990).
\bibitem{cao}
 N. Cao {\em et al.}, Phys. Rev. B {\bf 53}, 2601 (1996).
\bibitem{okamura}
 H. Okamura {\em et al.}, Phys. Rev. B {\bf 62}, R13265 (2000). 
\bibitem{rozenberg}
 M. J. Rozenberg, G. Kotliar, and H. Kajueter, Phys. Rev. B {\bf 54}, 8452 (1996).
\bibitem{jallen2} 
  J.W. Allen, and R. Martin, Phys. Rev. Lett. {\bf 49}, 1106 (1982) . 
\bibitem{gs}  
  O. Gunnarsson, and K. Sch\"onhammer, Phys. Rev. B {\bf 28}, 4315 (1983); {\bf 31}, 4815 (1985).
\bibitem{johansson} 
  B. Johansson, Philos. Mag. {\bf 30}, 469 (1974).
\bibitem{svane}
  J. Laegsgaard, and A. Svane, Phys. Rev. B {\bf 59}, 3450 (1999). 
\end{references}
\end{document}